\documentclass[onecolumn,superscriptaddress,10pt,nofootinbib]{revtex4}
\usepackage{graphicx}  
\usepackage{amsmath}   
\usepackage{amssymb}   
\usepackage{bm} 
\usepackage{dcolumn}
\usepackage{color}
\usepackage{mathrsfs}
\usepackage{amsfonts}
\usepackage{varioref}
\RequirePackage[colorlinks,citecolor=blue,urlcolor=magenta,linkcolor=blue]{hyperref}

\def\EH{Einstein-Hilbert }
\def\LL{Lanczos-Lovelock }

\def\gr{general relativity}

\labelformat{section}{Section #1} 
\labelformat{subsection}{Section #1} 
\labelformat{subsubsection}{Section #1}
\labelformat{subsubsubsection}{Section #1}
\labelformat{equation}{Eq.~(#1)} 
\labelformat{figure}{Fig.~#1} 
\labelformat{subfigure}{Fig.~\thefigure#1} 
\labelformat{table}{Tab.~#1} 
\labelformat{appendix}{Appendix #1}
\begin{document}
\title{ $1/r$ potential in higher dimensions}

\author{Sumanta Chakraborty}
\email{sumantac.physics@gmail.com}
\affiliation{Department of Theoretical Physics, Indian Association for the Cultivation of Science, Kolkata 700032, India}
\affiliation{IUCAA, Post Bag 4, Ganeshkhind, Pune University Campus, Pune 411 007, India}
\author{Naresh Dadhich}
\email{nkd@iucaa.in}
\affiliation{IUCAA, Post Bag 4, Ganeshkhind, Pune University Campus, Pune 411 007, India}
\affiliation{Center for Theoretical Physics, Jamia Millia Islamia, New Delhi 110025, India}


\begin{abstract}
In Einstein gravity, gravitational potential goes as $1/r^{d-3}$ in $d$ non-compactified spacetime dimensions, which assumes the familiar $1/r$ form in four dimensions. On the other hand, it goes as $1/r^{\alpha}$, with $\alpha=(d-2m-1)/m$, in pure Lovelock gravity involving only one $m$th order term of the Lovelock polynomial in the gravitational action. The latter offers a novel possibility of having $1/r$ potential for the non-compactified dimension spectrum given by $d=3m+1$. Thus it turns out that in the two prototype gravitational settings of isolated objects, like black holes and the universe as a whole --- cosmological models, the  Einstein gravity in four and $m$th order pure Lovelock gravity in $3m+1$ dimensions behave in a similar fashion as far as gravitational interactions are considered. However propagation of gravitational waves (or the number of degrees of freedom) does indeed serve as a discriminator because it has two polarizations only in four dimensions.
\end{abstract}
\maketitle
\section{Introduction}

For any metric theory of gravity what is required is construction of a divergence free second rank symmetric tensor from the metric and curvature alone. It can be achieved in two possible ways: (a) varying the action, which is an invariant quantity constructed from Riemann curvature and the metric, (b) by making trace of Bianchi derivative of a properly defined ``Riemann tensor'' to be vanishing. For Einstein gravity, variation of the \EH action with respect to metric tensor or trace of Bianchi identity of Riemann tensor, both lead to Einstein's equations \cite{gravitation,Carroll:1997ar,Parattu:2015gga}. In order to obtain physically meaningful solutions with proper initial value problem ensuring unique evolution as well as absence of ghosts, it is imperative that equations of motion continue to remain second order as is the case for Einstein gravity. The key requirement is that it should remain so even when we go to higher orders in the Riemann tensor. This requirement uniquely singles out Lovelock 
polynomial action that always yields second order gravitational field equations \cite{Lovelock:1971yv,Kastor:2008xb,Padmanabhan:2013xyr}. Lovelock action is essentially a sum of dimensionally continued Euler densities, being of order $m$ ($\geq 0$) in curvature, such that $m=1$ gives the Einstein-Hilbert Lagrangian, $m=2$ provides the Gauss-Bonnet Lagrangian and so on. Alternatively, it is possible to define $m$th order Lovelock ``Riemann'' tensor \cite{Dadhich:2008df,Kastor:2012se,Camanho:2015hea} and trace of its Bianchi derivative yields a divergence free second rank symmetric tensor, an analogue of Einstein tensor, as obtained from variation of $m$th order Lovelock action. 

\LL gravity is important for various reasons. It is generally believed that \EH action is an effective action, valid at small enough energy (or large length) scales. At high energy, near the Planck scale gravity cannot be described by \EH action alone but should be supplemented by higher curvature terms\footnote{Note that, requiring the field equations to be of second order uniquely singles out \LL Lagrangian from all other possibilities.}. For example, supersymmetric string theory \emph{exactly} reproduces Gauss-Bonnet term as quadratic correction to \EH action \cite{Boulware:1985wk}. Further thermodynamic interpretation of Einstein's equations generalize in a straightforward and natural \emph{but} non-trivial manner to \LL gravity. The same is true for other correspondences between thermodynamics and gravity in the context of \gr\ as well \cite{Chakraborty:2015wma,Chakraborty:2014joa,Chakraborty:2014rga,Chakraborty:2015hna}. Finally pure Lovelock theories, i.e., a single term in the Lovelock polynomial, 
exhibit very interesting features. These include --- (a) there is a close connection between pure Lovelock and dimensionally continued black holes \cite{Dadhich:2012ma,Garraffo:2008hu}, (b) gravity is kinematic in all critical $d=2m+1$ dimensions, i.e., vacuum is pure Lovelock flat \cite{Dadhich:2012cv}, (c) bound orbits exist for a given $m$ in all $2m+1<d<4m+1$ dimensions, in contrast, for Einstein gravity they do so only in four dimensions \cite{Dadhich:2013moa} and finally (d) equipartition of gravitational and non-gravitational energy defines location of black hole horizon \cite{Chakraborty:2015kva}. 

There have been various other interesting results in the context of cosmology within the Lovelock gravity, in particular the Einstein-Gauss-Bonnet gravity. For example, the scheme of dynamical compactification in higher curvature gravity introduced in \cite{MuellerHoissen:1985mm} has been applied to Einstein-Gauss-Bonnet gravity in higher dimensions to obtain exact solutions in various cosmological models \cite{Pavluchenko:2009wn,Maeda:2014nua,Canfora:2013xsa,Alexeyev:2000eb,Elizalde:2006ub}. The dynamical compactification scheme has also been applied in numerical investigations of cosmological scenarios in
\cite{Canfora:2014iga,Canfora:2008iu}. In particular it has been observed in \cite{Canfora:2016umq} that one can have standard Friedmann dynamics even though the gravity theory is that of Einstein-Gauss-Bonnet. It has also been noticed that in higher curvature gravity in higher dimensions the approach to big bang singularity has some discerning features in comparison to four dimensional gravity \cite{Deruelle:1989he}. However all the above works had Einstein-Gauss-Bonnet gravity as the background stage, also the results derived depend crucially on the compactification scheme while here we are concerned with a single term in the Lovelock polynomial and without any compactification scheme.

To these interesting features alluded above we wish to add in this note yet another interesting property of pure Lovelock gravity. In Einstein gravity potential goes as $1/r^{d-3}$ that takes the familiar $1/r$ form only in four dimensions and none else. In contrast, for pure Lovelock it goes as $1/r^\alpha$ with $\alpha=(d-2m-1)/m$ \cite{Dadhich:2012ma}, and hence there exists dimension spectrum $d=3m+1$ for $1/r$ potential. It should be emphasized that we are interested in exact solutions of gravitational field equations, not just Newtonian limit and hence must be contrasted with earlier approaches in \cite{ArkaniHamed:1998rs,ArkaniHamed:1999hk}. It is also possible to arrive at a similar situation in the weak field if the higher dimensional spacetime is assumed to be fractal in nature. However as demonstrated in \cite{Rami_CTP}, in order to arrive at a $1/r$ gravitational potential it was necessary to work in five dimensions with some particular choices related to the fractal structure of 
the spacetime (see also \cite{Tarasov_2015,Tarasov:2006bc,Tarasov:2014gua,Stillinger_1977}). However, the scenario we are interested in is completely different conceptually, as the spacetime manifold in consideration does not posses any fractal structure and the solution is derived from the fully non-linear theory in a series of spacetime dimensions. In our case the existence of $1/r$ potential is solely due to modified gravity theory rather than modified structure of spacetime manifold.

Therefore, the above feature can happen only in pure Lovelock gravity and hence this is also a nice distinguishing property for pure Lovelock theories. What it means is that, from a purely gravitational standpoint there is no way to fathom from solar system observations whether it is four dimensional Einstein or seven dimensional pure Gauss-Bonnet or in general $(3m+1)$-dimensional pure $m$th order Lovelock gravity. Further, it also turns out, that the same result applies to cosmological scenarios as well. In particular, in this work we have demonstrated that not only for dust but also for a linear equation of state parameter, which is being regularly considered in standard cosmology, the behaviour of the universe (as far as the gravitational sector is concerned) in pure Lovelock theories in $d=3m+1$ is indistinguishable from that of \gr\ in four dimensions, except for propagation of gravitational waves which would have two polarizations only in four dimensions and none else. 

Hence from purely gravitational standpoint, one cannot distinguish between any member of the $m$th order Lovelock in $d=3m+1$ dimensions. The discerning features do however come from two counts: (a) propagation of gravitational wave, and (b) introduction of matter fields. It turns out that gravitational wave can have two polarizations only in four dimensions. On the other hand, for existence of atoms, it is necessary that Maxwell field should admit bound orbits around a static charge which again can happen only in four dimensions. Thus these two features root us firmly to four dimensions which is in consonance with the standard paradigm of higher dimensional theories, where it is envisaged that all matter fields remain confined to the usual four dimension (termed as $3$-brane) while gravity is however free to propagate in higher dimensions. Therefore, at least, whether from a gravitational viewpoint we may as well be living in a higher dimensional spacetime rather than the usual four with gravity being 
described by pure Lovelock instead of Einstein gravity, is the question we wish to pose and wonder about in this note.  
\section{Lovelock Gravity}

In a $d$-dimensional spacetime, gravity in general, can be described by an action functional involving arbitrary scalar functions of metric and curvature, but not derivatives of curvature. In general, variation of this arbitrary Lagrangian would lead to an equation having fourth order derivatives of the metric. For it to be of second order, gravitational Lagrangian is constrained to be of the following Lovelock form,
\begin{equation}
L=\sum _{m}c_{m}L_{m}=\sum _{m}c_{m}\frac{1}{2^{m}}\delta ^{a_{1}b_{1}a_{2}b_{2}\ldots a_{m}b_{m}}_{c_{1}d_{1}c_{2}d_{2}\ldots c_{m}d_{m}}R^{c_{1}d_{1}}_{a_{1}b_{1}}R^{c_{2}d_{2}}_{a_{2}b_{2}}\ldots R^{c_{m}d_{m}}_{a_{m}b_{m}},
\end{equation}
where $\delta ^{pq\ldots}_{rs\ldots}$ stands for completely antisymmetric determinant tensor. The case $m=2$ is Gauss-Bonnet Lagrangian which is quadratic in curvature and reads as $L_{\rm GB}=(1/2)(R^{abcd}R_{abcd}-4R^{ab}R_{ab}+R^{2})$. Remarkably it also appears in the low energy effective theory of supersymmetric strings, as first pointed out in \cite{Boulware:1985wk}. Lovelock Lagrangian is a sum over $m$ where each term is a homogeneous polynomial in curvature and has a dimensionful coupling constant. Further, complete antisymmetry of the determinant tensor demands $d\geq 2m$, else it would vanish identically. Even for $d=2m$ \LL Lagrangian reduces to total derivative --- pure surface term. Lovelock Lagrangian $L_{m}$ is therefore non-trivial only in dimension $d\geq 2m+1$. Note that pure Lovelock gravity is kinematic in all critical odd $d=2m+1$ dimensions because $m$th order Riemann is entirely given in terms of corresponding Ricci, hence it has no non-trivial vacuum solution 
\cite{Dadhich:2012ma,Camanho:2015hea}. Thus non-trivial vacuum solutions only exist in dimensions $d\geq 2m+2$. Finally, variation of the Lagrangian with respect to metric variation, for pure Lovelock theories lead to the following second order equation,
\begin{equation}\label{pure_Eq_01}
{}^{(m)}E^{a}_{b}\equiv -\frac{1}{2^{m+1}}\delta ^{ac_{1}d_{1}\ldots c_{m}d_{m}}_{ba_{1}b_{1}\ldots a_{m}b_{m}}R^{a_{1}b_{1}}_{c_{1}d_{1}}\ldots R^{a_{m}b_{m}}_{c_{m}d_{m}}=8\pi T^{a}_{b}.
\end{equation}
Since there appears no derivatives of curvature, hence the equation is as expected of second order. Though not directly visible, second derivative too appears linearly; i.e., it is quasilinear, thereby ensuring unique evolution. 
\section{Black Holes}

Let us now consider pure Lovelock vacuum spacetime and write spherically symmetric metric in Schwarzschild gauge incorporating the null energy condition, as given by  
\begin{equation}\label{pure_ansatz}
ds^{2}=-f(r)dt^{2}+\frac{dr^{2}}{f(r)}+r^{2}d\Omega_{d-2}^{2},
\end{equation}
where $f(r)=1+2\Phi(r)$ and $d\Omega _{d-2}^{2}$ is metric on $(d-2)$-sphere. Then the equation, ${}^{(m)}E^{t}_{t}=0$ could be trivially integrated as it takes the simple form $(r^{d-2m-1} \Phi^m)' =0$ where prime is derivative relative to $r$, and so we obtain 
\begin{equation}
\Phi (r)=-\frac{M}{r^\alpha} , \, \, \, \alpha=(d-2m-1)/m,
\end{equation}
where $M$ is a constant of integration which identifies to $m$th root of the ADM mass \cite{Kastor:2008xb}. In general for static spherically symmetric metric in Schwarzschild gauge, the equation ${}^{(m)}E^{t}_{t}=0$ is the first integral of ${}^{(m)}E^{\theta}_{\theta}=0$, and so we have complete vacuum solution for pure Lovelock static black hole \cite{Dadhich:2012ma}.  

Clearly there is no way to distinguish between four dimensional Schwarzschild black hole and $(3m+1)$ dimensional $m$th order pure Lovelock black hole because the gravitational potential goes as $1/r$ for all of them. It is worth emphasizing that it is not only the Newtonian limit that behaves as $1/r$, but the full non-linear Lovelock field equation that has the following exact solution,
\begin{equation}
ds^{2}_{d=3m+1}=-\left(1-\frac{2M}{r}\right)dt^{2}+\left(1-\frac{2M}{r}\right)^{-1}dr^{2}
+r^{2}d\Omega _{3m-1}^{2}
\end{equation}
such that on the four-dimensional hypersurface characterized by $\theta _{3}=\theta _{4}=\cdots =\theta _{3m-3}=\pi/2$, the spacetime metric is \emph{identical} to the Schwarzschild solution in four spacetime dimensions. Note that we have not employed any compactification scheme in order to arrive at this solution. Hence any observable tests in solar system physics cannot distinguish between $m$th order Lovelock gravity in $d=3m+1$ dimensions and Einstein gravity in four dimensions.

As a matter of fact dimensional spectrum is given by $d=(2+\alpha)m+1$ corresponding to a given value of $\alpha$. The choice $\alpha=2$ at the first level corresponds to five dimensional Einstein with dimensional spectrum $d=4m+1$ while $\alpha=1/2$ for six dimensional Gauss-Bonnet with dimensional spectrum $d=5m/2+1$. Also note that since horizon structure and potential are the same, all properties associated with horizon, e.g., Hawking radiation, should remain the same. Another test that can distinguish these two black holes may be \emph{black hole entropy}. But \emph{remarkably} in the case under consideration, entropy density of the black hole goes as product of $(m-1)$ curvature, each of which scales as $r^{-3(m-1)}$ for $d=3m+1$, while area element scales as $r^{3m-1}$ and hence entropy would always scale as $r_h^{2}$, where $r_{h}$ is horizon radius irrespective of Lovelock order $m$. For pure Lovelock static black holes, it has been shown that thermodynamical parameters, temperature and entropy bear 
the same relation to horizon radius in all critical odd and even $d=2m+1, 2m+2$ dimensions \cite{Dadhich:2012eg}.  

The next question arises, would black hole, like Schwarzschild in four dimension, be stable in higher dimensions? In stability analysis one decomposes metric perturbations into three types, tensor, vector and scalar perturbations \cite{Regge:1957td,Vishveshwara:1970cc,Ishibashi:2003ap}. Stability analysis of Lovelock gravity black holes under scalar, vector and tensor perturbations was carried out in \cite{Takahashi:2009xh}, which was further particularized to pure Lovelock black holes in \cite{Gannouji:2013eka}. Under metric perturbation, Riemann tensor and hence all geometric constructs will be perturbed. Using which one can obtain a master differential equation for perturbation, which is like Schr\"{o}dinger equation with a potential. Only when  corresponding differential operator is a positive (or zero) self-adjoint operator, solutions will be perturbatively stable if and only if there are no negative eigenvalues. This in turn leads to the following condition 
\begin{equation}
\frac{d-3m-1}{mr}\geq 0.
\end{equation}
Clearly it would be stable for $d\geq3m+1$ which means $\alpha =(d-2m-1)/m\geq1$. Also note that the same conclusion was drawn from a different perspective for stability of small black holes \cite{Camanho:2013kfa} in generic Lovelock theories. This is another interesting feature of our work. The above result explicitly shows that black hole for Einstein gravity in four or higher dimensions is always stable because $\alpha\geq1$, while it is unstable for pure Lovelock gravity in all the critical even $d=2m+2$ dimensions \footnote{Note that $d=3m+1$ is not the critical dimension for $m$-th order pure Lovelock.} for which $\alpha = 1/m <1$ \cite{Gannouji:2013eka}. Thus the determining factor is $\alpha\geq1$. Since it is stable under tensor perturbations and hence it would be so for vector perturbations as well. 

The final question corresponds to the weak field limit of gravity, which is generally obtained by assuming the metric to be a perturbation over a flat background, i.e., $g_{ab}=\eta _{ab}+h_{ab}$ with $|h_{ab}|\ll 1$. In general Lovelock polynomial the Einstein-Hilbert term is always present, thus the weak field limit will contain linear order terms in the perturbation. However in the case of pure Lovelock theories only a single term in the Lovelock polynomial is present, there is no term linear in $h_{ab}$ because curvature tensor for the background geometry vanishes. For ``m''-th order pure Lovelock theories the weak field limit would correspond to a differential equation having m powers of the perturbation $h_{ab}$. For example, in the case of pure Gauss-Bonnet gravity the gravitational field equation in the weak field limit would involve terms like $(1/4)\square \bar{h}_{ab}\square \bar{h}^{ab}$, $\partial _{a}\partial _{b}\bar{h}_{cd}\partial ^{m}\partial ^{b}\bar{h}^{cd}$ and so on, where $\bar{h}_{ab}
=h_{ab}-(1/2)\eta _{ab}h$ \cite{rado}. Hence the Newtonian limit is not reproduced.

This suggests that for Lovelock gravity, perhaps one should consider perturbation around constant, not zero, curvature spacetime; i.e., dS/AdS which would yield non-trivial equation. For that purpose we write $g_{ab}=\bar{g}_{ab}+h_{ab}$, where $\bar{g}_{ab}$ is the background dS/AdS spacetime and $h_{ab}$ is the perturbation. Then the curvature tensor to leading order in the perturbation leads to, 
\begin{equation}
R^{ab}_{cd}=M\delta ^{a}_{c}\delta ^{b}_{d}+\frac{1}{2}\nabla _{c}\left(-\nabla ^{a}h^{b}_{d}+\nabla _{d}h^{ab}+\nabla ^{b}h^{a}_{d}\right)-\left(c\leftrightarrow d\right)
\end{equation}
where $M$ is proportional to the Ricci scalar and hence to the cosmological constant. Use of this expansion leads to the following expression of the field equation expanded upto first order in the perturbation from \ref{pure_Eq_01} as,
\begin{eqnarray}
-\frac{M^{m-1}}{2}&\times&\delta ^{aa_{1}b_{1}\ldots a_{m}b_{m}}_{bc_{1}d_{1}\ldots c_{m}d_{m}}\delta ^{c_{2}}_{a_{2}}\delta ^{d_{2}}_{b_{2}}\ldots \delta ^{c_{m}}_{a_{m}}\delta ^{d_{m}}_{b_{m}}\Bigg[\delta ^{c_{1}}_{a_{1}}\delta ^{d_{1}}_{b_{1}}
\nonumber
\\
&+&\frac{1}{2}\nabla _{a_{1}}\left(-\nabla ^{c_{1}}h^{d_{1}}_{b_{1}}+\nabla _{b_{1}}h^{c_{1}d_{1}}+\nabla ^{d_{1}}h^{c_{1}}_{b_{1}}\right)\Bigg]=8\pi T^{a}_{b}
\end{eqnarray}
One can evaluate the left hand side explicitly. The first term will lead to the standard cosmological constant contribution, which as subtracted away will lead to the following evolution equation of the perturbation in vacuum spacetime 
\begin{equation}
\square \bar{h}_{ab}=0
\end{equation}
where $\bar{h}_{ab}=h_{ab}-(1/2)\bar{g}_{ab}h$ and we have used the gauge condition $\nabla _{a}\bar{h}^{ab}=0$. The above expression is essentially the same as the one obtained from Einstein gravity in the weak field limit. Thus in presence of cosmological constant the weak field limit of Einstein gravity can be reproduced.  

This clearly shows that pure Lovelock gravity picks out constant curvature spacetime as background rather than the flat Minkowski. In this context it is interesting to note that on inclusion of $\Lambda$ we shall have $f(r)=1-(\Lambda r^{2m} + M/r^{d-2m-1})^{1/m}$ which for large $r$ goes to $f(r)= 1 - \bar\Lambda r^2 - 2\bar{M}/r^{d-3}$ \cite{Dadhich:2012ma}. This is the $d$-dimensional Einstein-dS solution. It is remarkable that even though there was no Einstein term in the equation of motion yet the solution goes over to Einstein-dS asymptotically. 

At this stage, let us summarize the key results that we have arrived at in the context of black holes in pure Lovelock theories. First of all, it is possible to have \emph{exact} Schwarzschild solution on a four dimensional hypersurface for $m$th order pure Lovelock gravity in $d=3m+1$ dimensions. Secondly, the semi-classical properties of the spacetime, e.g., Hawking radiation and black hole entropy for this $(3m+1)$ dimensional Lovelock black hole remain indistinguishable from the four dimensional Einstein gravity. Moreover it turns out that the black hole in $(3m+1)$ dimension is indeed stable alike the Schwarzschild solution in four dimensions. Finally correct weak field limit in these theories can also be obtained when expanded around de Sitter or anti-de Sitter solution.

Since the weak field limit of pure Lovelock gravity is identical to that of Einstein, it would be interesting to understand the propagation of gravitational waves in higher dimensional pure Lovelock background, in particular the structure of the corresponding quadrupole formula will be worth pursuing. Even without going into details, which we keep for the future, one can provide some general comments. Note that presence of cosmological constant is mandatory in pure Lovelock theories in order to arrive at the correct weak field limit, while the quadrupole moment formula is derived in \gr\ without a cosmological constant. Recently, there have been renewed interest in the gravitational wave propagation in de Sitter spacetime, whose formulation differs quite  significantly from the standard scenario and results in a modified quadrupole moment formula \cite{Ashtekar:2015lxa,Ashtekar:2015ooa,Ashtekar:2014zfa}. Thus even in the context of pure Lovelock theories the quadrupole formula will be different from the \gr\ 
counterpart. However it would be interesting to see does it bear any resemblance with the quadrupole formula for de Sitter spacetime or not?  In particular whether the dimensional spectrum also holds in this context?
\section{FLRW Cosmology}

From black hole we now come to cosmology and ask the question, would the standard FLRW cosmology remain the same for the dimensional spectrum $d=3m+1$? That means this feature is not only restricted to vacuum solutions but is true for cosmology as well. Thus the standard FLRW metric for homogeneous and isotropic universe reads,
\begin{equation}
ds^{2}=-dt^{2}+a^{2}(t)\left[dr^{2}+r^{2}d\Omega _{d-2}^{2}\right].
\end{equation}
From Eq. (2) we obtain for Hubble parameter, $H=\dot{a}/a$ as, 
\begin{equation}\label{pure_cosmo_01}
H^{2m}=\frac{(d-2m-1)!}{(d-1)!}16\pi G\rho,
\end{equation}
and for isotropic pressure 
\begin{eqnarray}\label{pure_cosmo_02}
\frac{1}{2}\frac{(d-2)!}{(d-2m-1)!}H^{2m-2}&\times&\left[2m\frac{\ddot{a}}{a}+(d-2m-1)H^{2}\right]=-8\pi Gp
\end{eqnarray}
Deriving the conservation equation from \ref{pure_cosmo_01} and \ref{pure_cosmo_02} is straightforward, one needs to take time derivative of \ref{pure_cosmo_01} and then use \ref{pure_cosmo_02} to eliminate $\dot{H}$ term, resulting in $\dot{\rho}+(d-1)(\rho +p)H=0$. Assuming an equation of state $p=\omega \rho$ with a constant parameter $\omega$ we obtain time evolution of scale factor as $a(t)\sim t^{2m/[(d-1)(1+w)]}$. It is immediate that for $d=3m+1$, one obtains $a(t)\sim t^{2/[3(1+w)]}$, identical to the standard general relativistic cosmology in these contexts. 

Note that in the above cosmological scenario the equation of state parameter, $\omega$ can have any value including $1/3, 0, -1$ respectively for early time radiation, non-relativistic dust fluid and the cosmological constant -- the dark energy. Thus it fully accords to standard cosmological model which does however have open problems of dark matter and dark energy. For addressing these problems, one either seeks modification of gravity theory or some exotic matter field. Our aim was not to address these questions instead to point out that there can exist exactly the same cosmological scenario in higher dimensions in an appropriate gravity theory.

However all this would have been fine, had matter fields could have also existed in higher dimensions, which is however not the case. Hence even though at face value the cosmological solutions are identical to their general relativistic counterparts, inclusion of matter (in particular, radiation) acts as the discriminator. That is, our Universe may be four dimensional with Einstein gravity or seven dimensional with Gauss-Bonnet gravity, and there is no way to break this dilemma from a pure gravitational standpoint. But, introduction of matter clearly breaks it in favour of 4-dimensional Einstein gravity.
\section{Discussion}

There is a very strong and compelling observational evidence for $1/r$ potential in four dimensions in the framework of general relativity which is a metric theory of gravity. It is curious to ask, could gravitational potential have the same behaviour in any other dimension in some other metric theory? It is remarkable that the answer is yes, and it is uniquely the pure Lovelock theory that offers a spectrum of dimensions, $d=3m+1$ for which potential is indeed $1/r$. This means all gravitational observations, except the gravitational waves, involving stellar objects and black holes will not be able to distinguish between any two members of this spectrum. Not only that, cosmological dynamics is also driven by the same inverse square law and hence it is expected that should also remain the same. And that indeed is the case as we have just shown above. Below we point out the main outcomes of this work:
\begin{itemize}
  
\item We have explicitly demonstrated that the two most fundamental problems for a gravitational theory, namely compact objects and cosmology might have the same dynamics for this dimensional spectrum from a purely gravitational standpoint. This is a very important conclusion that simply follows from pure Lovelock gravity.

\item We have also demonstrated that all the properties associated with the black hole horizon, e.g., Hawking radiation, remains identical with the four-dimensional counterpart. Moreover, the horizon entropy would always scale as $r_h^{2}$, where $r_{h}$ is the horizon radius, irrespective of the Lovelock order $m$ (also see, \cite{Dadhich:2012eg}). This is another intriguing fact associated with the pure Lovelock theories.

\item We have also shown that, when perturbed with respect to constant curvature background, the weak field limit of Einstein gravity can be reproduced. It also follows that pure Lovelock static black hole solution with $\Lambda$, asymptotically goes over to  $d$-dimensional Einstein-dS solution \cite{Dadhich:2012ma}. It is intriguing that even though there was no Einstein term present in the equation of motion, yet the solution goes over to Einstein-dS asymptotically.

\end{itemize}
In particular, purely based on gravitational observations it would be difficult to decipher whether it is the usual four dimensional spacetime with Einstein gravity or ten dimensional spacetime with pure third order Lovelock gravity. 

Besides the question of existence of matter in higher dimensions, there is yet another gravitational feature that may also turn out to be a discriminator. It is the propagation of gravitational wave and its number of polarizations and degrees of freedom. It had been argued that number of degrees of freedom in a $d$ dimensional spacetime remain the same as given by $d(d-3)/2$ for Einstein as well as Lovelock theory \cite{Teitelboim:1987zz,Deser:2011zk}, and it has recently been further reinforced for pure Lovelock gravity \cite{Dadhich:2015ivt}. In that case two polarizations for gravitational wave could only occur in four dimension and none else. Let us see how count for degrees of freedom works in a $d$ dimensional spacetime by separating out the usual graviton degrees, gauge vector, and scalar degrees in the light of Kaluza-Klein decomposition. Let us envisage the metric in $d$ dimensions with parametrization: $g_{AB} =({g_{ab}, g_{ai}, g_{ij}})$ where $A,B=0,\ldots,(d-1); a,b=0,1,2,3; i,j=4,
5,\ldots,(d-1)$. Then there is the usual graviton $g_{ab}$ with two degrees, $(d-4)$ gauge vector fields with each two degrees and $(d-4)(d-3)/2$ scalars. One can easily check that they add up to give the correct number of degrees of freedom $d(d-3)/2$. Thus extra degrees of freedom would be $(d+1)(d-4)/2$, which becomes, $3(m-1)(3m+2)/2$ for $d=3m+1$.

For seven dimensional Gauss-Bonnet spacetime, extra degrees of freedom include, three gauge vector fields, each having two degrees and six scalars. So there would appear extra vector and scalar fields which could possibly distinguish between the usual four dimensional Einstein and seven dimensional Gauss-Bonnet spacetime. However for a static source in seven dimensions, the metric is orthogonal and thus eliminating all the gauge fields and three scalar fields, and the remaining three scalars will be washed out on dimensional reduction, $\theta_i = \pi/2$. Despite all this a comprehensive field theoretic analysis has to be undertaken on the lines of Kaluza-Klein theory \cite{Appelquist:1983vs} before anything definitive could be said. However, it seems that propagation of gravitational waves or inclusion of matter in the theory will vote for four-dimensional Schwarzschild spacetime alone, unless existence of matter in higher dimensions is assumed.

We thus conclude that even though it is difficult gravitationally to distinguish between any two members of the dimensional spectrum, except for the number of degrees of freedom, raising the question, ``Could we have as well lived in higher dimensions?", perhaps yes, if matter could have existed there. It possibly points to the fact that gravity is abundant in complexity and richness of structure that one is never disappointed as one probes deeper and wider. The present question offers an excellent example of this richness as it coaxes one to wonder, could the universe been higher dimensional or not? This is an important question that would have broad scientific as well as philosophical fallout.  
\section*{Acknowledgements}

Research of S.C. is funded by the SERB-NPDF grant (PDF/2016/001589) from SERB, Government of India. He also thanks Kinjalk Lochan for fruitful discussions. N.D. thanks Xian Camanho for a very enlightening discussion on degrees of freedom and also acknowledges warm hospitality of Albert Einstein Institute where the manuscript was finalized. 
\bibliography{Gravity_1_full,Gravity_2_full,Gravity_3_partial,Brane,My_References}

\providecommand{\href}[2]{#2}\begingroup\raggedright\begin{thebibliography}{10}

\bibitem{gravitation}
T.Padmanabhan, {\em {Gravitation: Foundations and Frontiers}}.
\newblock Cambridge University Press, Cambridge, UK, 2010.

\bibitem{Carroll:1997ar}
S.~M. Carroll, ``{Lecture notes on general relativity},''
\href{http://arxiv.org/abs/gr-qc/9712019}{{\ttfamily arXiv:gr-qc/9712019
  [gr-qc]}}.

\bibitem{Parattu:2015gga}
K.~Parattu, S.~Chakraborty, B.~R. Majhi, and T.~Padmanabhan, ``{A Boundary Term
  for the Gravitational Action with Null Boundaries},''
  \href{http://dx.doi.org/10.1007/s10714-016-2093-7}{{\em Gen. Rel. Grav.}
  {\bfseries 48} no.~7, (2016) 94},
\href{http://arxiv.org/abs/1501.01053}{{\ttfamily arXiv:1501.01053 [gr-qc]}}.

\bibitem{Lovelock:1971yv}
D.~Lovelock, ``{The Einstein tensor and its generalizations},''
\href{http://dx.doi.org/10.1063/1.1665613}{{\em J. Math. Phys.} {\bfseries 12}
  (1971) 498--501}.

\bibitem{Kastor:2008xb}
D.~Kastor, ``{Komar Integrals in Higher (and Lower) Derivative Gravity},''
  \href{http://dx.doi.org/10.1088/0264-9381/25/17/175007}{{\em
  Class.Quant.Grav.} {\bfseries 25} (2008) 175007},
\href{http://arxiv.org/abs/0804.1832}{{\ttfamily arXiv:0804.1832 [hep-th]}}.

\bibitem{Padmanabhan:2013xyr}
T.~Padmanabhan and D.~Kothawala, ``{Lanczos-Lovelock models of gravity},''
  \href{http://dx.doi.org/10.1016/j.physrep.2013.05.007}{{\em Phys.Rept.}
  {\bfseries 531} (2013) 115--171},
\href{http://arxiv.org/abs/1302.2151}{{\ttfamily arXiv:1302.2151 [gr-qc]}}.

\bibitem{Dadhich:2008df}
N.~Dadhich, ``{Characterization of the Lovelock gravity by Bianchi
  derivative},'' \href{http://dx.doi.org/10.1007/s12043-010-0080-1}{{\em
  Pramana} {\bfseries 74} (2010) 875--882},
\href{http://arxiv.org/abs/0802.3034}{{\ttfamily arXiv:0802.3034 [gr-qc]}}.

\bibitem{Kastor:2012se}
D.~Kastor, ``{The Riemann-Lovelock Curvature Tensor},''
  \href{http://dx.doi.org/10.1088/0264-9381/29/15/155007}{{\em Class. Quant.
  Grav.} {\bfseries 29} (2012) 155007},
\href{http://arxiv.org/abs/1202.5287}{{\ttfamily arXiv:1202.5287 [hep-th]}}.

\bibitem{Camanho:2015hea}
X.~O. Camanho and N.~Dadhich, ``{On Lovelock analogs of the Riemann tensor},''
  \href{http://dx.doi.org/10.1140/epjc/s10052-016-3891-5}{{\em Eur. Phys. J.}
  {\bfseries C76} no.~3, (2016) 149},
\href{http://arxiv.org/abs/1503.02889}{{\ttfamily arXiv:1503.02889 [gr-qc]}}.

\bibitem{Boulware:1985wk}
D.~G. Boulware and S.~Deser, ``{String Generated Gravity Models},''
\href{http://dx.doi.org/10.1103/PhysRevLett.55.2656}{{\em Phys. Rev. Lett.}
  {\bfseries 55} (1985) 2656}.

\bibitem{Chakraborty:2015wma}
S.~Chakraborty, ``{Lanczos-Lovelock gravity from a thermodynamic
  perspective},'' \href{http://dx.doi.org/10.1007/JHEP08(2015)029}{{\em JHEP}
  {\bfseries 08} (2015) 029},
\href{http://arxiv.org/abs/1505.07272}{{\ttfamily arXiv:1505.07272 [gr-qc]}}.

\bibitem{Chakraborty:2014joa}
S.~Chakraborty and T.~Padmanabhan, ``{Geometrical variables with direct
  thermodynamic significance in Lanczos-Lovelock gravity},''
  \href{http://dx.doi.org/10.1103/PhysRevD.90.084021}{{\em Phys.Rev.}
  {\bfseries D90} no.~8, (2014) 084021},
\href{http://arxiv.org/abs/1408.4791}{{\ttfamily arXiv:1408.4791 [gr-qc]}}.

\bibitem{Chakraborty:2014rga}
S.~Chakraborty and T.~Padmanabhan, ``{Evolution of Spacetime arises due to the
  departure from Holographic Equipartition in all Lanczos-Lovelock Theories of
  Gravity},'' \href{http://dx.doi.org/10.1103/PhysRevD.90.124017}{{\em
  Phys.Rev.} {\bfseries D90} no.~12, (2014) 124017},
\href{http://arxiv.org/abs/1408.4679}{{\ttfamily arXiv:1408.4679 [gr-qc]}}.

\bibitem{Chakraborty:2015hna}
S.~Chakraborty and T.~Padmanabhan, ``{Thermodynamical interpretation of the
  geometrical variables associated with null surfaces},''
  \href{http://dx.doi.org/10.1103/PhysRevD.92.104011}{{\em Phys. Rev.}
  {\bfseries D92} no.~10, (2015) 104011},
\href{http://arxiv.org/abs/1508.04060}{{\ttfamily arXiv:1508.04060 [gr-qc]}}.

\bibitem{Dadhich:2012ma}
N.~Dadhich, J.~M. Pons, and K.~Prabhu, ``{On the static Lovelock black
  holes},'' \href{http://dx.doi.org/10.1007/s10714-013-1514-0}{{\em
  Gen.Rel.Grav.} {\bfseries 45} (2013) 1131--1144},
\href{http://arxiv.org/abs/1201.4994}{{\ttfamily arXiv:1201.4994 [gr-qc]}}.

\bibitem{Garraffo:2008hu}
C.~Garraffo and G.~Giribet, ``{The Lovelock Black Holes},''
  \href{http://dx.doi.org/10.1142/S0217732308027497}{{\em Mod. Phys. Lett.}
  {\bfseries A23} (2008) 1801--1818},
\href{http://arxiv.org/abs/0805.3575}{{\ttfamily arXiv:0805.3575 [gr-qc]}}.

\bibitem{Dadhich:2012cv}
N.~Dadhich, S.~G. Ghosh, and S.~Jhingan, ``{The Lovelock gravity in the
  critical spacetime dimension},''
  \href{http://dx.doi.org/10.1016/j.physletb.2012.03.084}{{\em Phys.Lett.}
  {\bfseries B711} (2012) 196--198},
\href{http://arxiv.org/abs/1202.4575}{{\ttfamily arXiv:1202.4575 [gr-qc]}}.

\bibitem{Dadhich:2013moa}
N.~Dadhich, S.~G. Ghosh, and S.~Jhingan, ``{Bound orbits and gravitational
  theory},'' \href{http://dx.doi.org/10.1103/PhysRevD.88.124040}{{\em Phys.
  Rev.} {\bfseries D88} no.~12, (2013) 124040},
\href{http://arxiv.org/abs/1308.4770}{{\ttfamily arXiv:1308.4770 [gr-qc]}}.

\bibitem{Chakraborty:2015kva}
S.~Chakraborty and N.~Dadhich, ``{Brown-York quasilocal energy in
  Lanczos-Lovelock gravity and black hole horizons},''
  \href{http://dx.doi.org/10.1007/JHEP12(2015)003}{{\em JHEP} {\bfseries 12}
  (2015) 003},
\href{http://arxiv.org/abs/1509.02156}{{\ttfamily arXiv:1509.02156 [gr-qc]}}.

\bibitem{MuellerHoissen:1985mm}
F.~Mueller-Hoissen, ``{Spontaneous Compactification With Quadratic and Cubic
  Curvature Terms},''
\href{http://dx.doi.org/10.1016/0370-2693(85)90202-3}{{\em Phys. Lett.}
  {\bfseries B163} (1985) 106--110}.

\bibitem{Pavluchenko:2009wn}
S.~A. Pavluchenko, ``{General features of Bianchi-I cosmological models in
  Lovelock gravity},'' \href{http://dx.doi.org/10.1103/PhysRevD.80.107501}{{\em
  Phys. Rev.} {\bfseries D80} (2009) 107501},
\href{http://arxiv.org/abs/0906.0141}{{\ttfamily arXiv:0906.0141 [gr-qc]}}.

\bibitem{Maeda:2014nua}
K.-i. Maeda and N.~Ohta, ``{Cosmic acceleration with a negative cosmological
  constant in higher dimensions},''
  \href{http://dx.doi.org/10.1007/JHEP06(2014)095}{{\em JHEP} {\bfseries 06}
  (2014) 095},
\href{http://arxiv.org/abs/1404.0561}{{\ttfamily arXiv:1404.0561 [hep-th]}}.

\bibitem{Canfora:2013xsa}
F.~Canfora, A.~Giacomini, and S.~A. Pavluchenko, ``{Dynamical compactification
  in Einstein-Gauss-Bonnet gravity from geometric frustration},''
  \href{http://dx.doi.org/10.1103/PhysRevD.88.064044}{{\em Phys. Rev.}
  {\bfseries D88} no.~6, (2013) 064044},
\href{http://arxiv.org/abs/1308.1896}{{\ttfamily arXiv:1308.1896 [gr-qc]}}.

\bibitem{Alexeyev:2000eb}
S.~Alexeyev, A.~Toporensky, and V.~Ustiansky, ``{The Nature of singularity in
  Bianchi I cosmological string gravity model with second order curvature
  corrections},'' \href{http://dx.doi.org/10.1016/S0370-2693(01)00556-1}{{\em
  Phys. Lett.} {\bfseries B509} (2001) 151--156},
\href{http://arxiv.org/abs/gr-qc/0009020}{{\ttfamily arXiv:gr-qc/0009020
  [gr-qc]}}.

\bibitem{Elizalde:2006ub}
E.~Elizalde, A.~N. Makarenko, V.~V. Obukhov, K.~E. Osetrin, and A.~E. Filippov,
  ``{Stationary vs. singular points in an accelerating FRW cosmology derived
  from six-dimensional Einstein-Gauss-Bonnet gravity},''
  \href{http://dx.doi.org/10.1016/j.physletb.2006.11.031}{{\em Phys. Lett.}
  {\bfseries B644} (2007) 1--6},
\href{http://arxiv.org/abs/hep-th/0611213}{{\ttfamily arXiv:hep-th/0611213
  [hep-th]}}.

\bibitem{Canfora:2014iga}
F.~Canfora, A.~Giacomini, and S.~A. Pavluchenko, ``{Cosmological dynamics in
  higher-dimensional Einstein-Gauss-Bonnet gravity},''
  \href{http://dx.doi.org/10.1007/s10714-014-1805-0}{{\em Gen. Rel. Grav.}
  {\bfseries 46} no.~10, (2014) 1805},
\href{http://arxiv.org/abs/1409.2637}{{\ttfamily arXiv:1409.2637 [gr-qc]}}.

\bibitem{Canfora:2008iu}
F.~Canfora, A.~Giacomini, R.~Troncoso, and S.~Willison, ``{General Relativity
  with small cosmological constant from spontaneous compactification of
  Lovelock theory in vacuum},''
  \href{http://dx.doi.org/10.1103/PhysRevD.80.044029}{{\em Phys. Rev.}
  {\bfseries D80} (2009) 044029},
\href{http://arxiv.org/abs/0812.4311}{{\ttfamily arXiv:0812.4311 [hep-th]}}.

\bibitem{Canfora:2016umq}
F.~Canfora, A.~Giacomini, S.~A. Pavluchenko, and A.~Toporensky, ``{Friedmann
  dynamics recovered from compactified Einstein-Gauss-Bonnet cosmology},''
\href{http://arxiv.org/abs/1605.00041}{{\ttfamily arXiv:1605.00041 [gr-qc]}}.

\bibitem{Deruelle:1989he}
N.~Deruelle, ``{On the Approach to the Cosmological Singularity in Quadratic
  Theories of Gravity: The Kasner Regimes},''
\href{http://dx.doi.org/10.1016/0550-3213(89)90294-0}{{\em Nucl. Phys.}
  {\bfseries B327} (1989) 253--266}.

\bibitem{ArkaniHamed:1998rs}
N.~Arkani-Hamed, S.~Dimopoulos, and G.~Dvali, ``{The Hierarchy problem and new
  dimensions at a millimeter},''
  \href{http://dx.doi.org/10.1016/S0370-2693(98)00466-3}{{\em Phys.Lett.}
  {\bfseries B429} (1998) 263--272},
\href{http://arxiv.org/abs/hep-ph/9803315}{{\ttfamily arXiv:hep-ph/9803315
  [hep-ph]}}.

\bibitem{ArkaniHamed:1999hk}
N.~Arkani-Hamed, S.~Dimopoulos, G.~R. Dvali, and N.~Kaloper, ``{Infinitely
  large new dimensions},''
  \href{http://dx.doi.org/10.1103/PhysRevLett.84.586}{{\em Phys. Rev. Lett.}
  {\bfseries 84} (2000) 586--589},
\href{http://arxiv.org/abs/hep-th/9907209}{{\ttfamily arXiv:hep-th/9907209
  [hep-th]}}.

\bibitem{Rami_CTP}
R.~A. El-Nabulsi, ``New metrics from a fractional gravitational field,'' {\em
  Communications in Theoretical Physics} {\bfseries 68} no.~3, (2017) 309.
  \url{http://stacks.iop.org/0253-6102/68/i=3/a=309}.

\bibitem{Tarasov_2015}
V.~E. {Tarasov}, ``{Vector calculus in non-integer dimensional space and its
  applications to fractal media},''
  \href{http://dx.doi.org/10.1016/j.cnsns.2014.05.025}{{\em Communications in
  Nonlinear Science and Numerical Simulations} {\bfseries 20} (Feb., 2015)
  360--374}, \href{http://arxiv.org/abs/1503.02022}{{\ttfamily arXiv:1503.02022
  [math-ph]}}.

\bibitem{Tarasov:2006bc}
V.~E. Tarasov, ``{Gravitational Field of Fractal Distribution of Particles},''
  \href{http://dx.doi.org/10.1007/s10569-005-1152-2}{{\em Celest. Mech. Dyn.
  Astron.} {\bfseries 94} (2006) 1--15},
\href{http://arxiv.org/abs/astro-ph/0604491}{{\ttfamily arXiv:astro-ph/0604491
  [astro-ph]}}.

\bibitem{Tarasov:2014gua}
V.~E. Tarasov, ``{Fractional Quantum Field Theory: From Lattice to
  Continuum},''
\href{http://dx.doi.org/10.1155/2014/957863}{{\em Adv. High Energy Phys.}
  {\bfseries 2014} (2014) 957863}.

\bibitem{Stillinger_1977}
F.~H. {Stillinger}, ``{Axiomatic basis for spaces with noninteger dimension},''
  \href{http://dx.doi.org/10.1063/1.523395}{{\em Journal of Mathematical
  Physics} {\bfseries 18} (June, 1977) 1224--1234}.

\bibitem{Dadhich:2012eg}
N.~Dadhich, J.~M. Pons, and K.~Prabhu, ``{Thermodynamical universality of the
  Lovelock black holes},''
  \href{http://dx.doi.org/10.1007/s10714-012-1416-6}{{\em Gen. Rel. Grav.}
  {\bfseries 44} (2012) 2595--2601},
\href{http://arxiv.org/abs/1110.0673}{{\ttfamily arXiv:1110.0673 [gr-qc]}}.

\bibitem{Regge:1957td}
T.~Regge and J.~A. Wheeler, ``{Stability of a Schwarzschild singularity},''
\href{http://dx.doi.org/10.1103/PhysRev.108.1063}{{\em Phys. Rev.} {\bfseries
  108} (1957) 1063--1069}.

\bibitem{Vishveshwara:1970cc}
C.~V. Vishveshwara, ``{Stability of the schwarzschild metric},''
\href{http://dx.doi.org/10.1103/PhysRevD.1.2870}{{\em Phys. Rev.} {\bfseries
  D1} (1970) 2870--2879}.

\bibitem{Ishibashi:2003ap}
A.~Ishibashi and H.~Kodama, ``{Stability of higher dimensional Schwarzschild
  black holes},'' \href{http://dx.doi.org/10.1143/PTP.110.901}{{\em Prog.
  Theor. Phys.} {\bfseries 110} (2003) 901--919},
\href{http://arxiv.org/abs/hep-th/0305185}{{\ttfamily arXiv:hep-th/0305185
  [hep-th]}}.

\bibitem{Takahashi:2009xh}
T.~Takahashi and J.~Soda, ``{Instability of Small Lovelock Black Holes in
  Even-dimensions},'' \href{http://dx.doi.org/10.1103/PhysRevD.80.104021}{{\em
  Phys. Rev.} {\bfseries D80} (2009) 104021},
\href{http://arxiv.org/abs/0907.0556}{{\ttfamily arXiv:0907.0556 [gr-qc]}}.

\bibitem{Gannouji:2013eka}
R.~Gannouji and N.~Dadhich, ``{Stability and existence analysis of static black
  holes in pure Lovelock theories},''
  \href{http://dx.doi.org/10.1088/0264-9381/31/16/165016}{{\em Class. Quant.
  Grav.} {\bfseries 31} (2014) 165016},
\href{http://arxiv.org/abs/1311.4543}{{\ttfamily arXiv:1311.4543 [gr-qc]}}.

\bibitem{Camanho:2013kfa}
X.~O. Camanho and J.~D. Edelstein, ``{Cosmic censorship in Lovelock theory},''
  \href{http://dx.doi.org/10.1007/JHEP11(2013)151}{{\em JHEP} {\bfseries 11}
  (2013) 151},
\href{http://arxiv.org/abs/1308.0304}{{\ttfamily arXiv:1308.0304 [hep-th]}}.

\bibitem{rado}
R.~Gannouji, ``Private communication,''.

\bibitem{Ashtekar:2015lxa}
A.~Ashtekar, B.~Bonga, and A.~Kesavan, ``{Asymptotics with a positive
  cosmological constant: III. The quadrupole formula},''
  \href{http://dx.doi.org/10.1103/PhysRevD.92.104032}{{\em Phys. Rev.}
  {\bfseries D92} no.~10, (2015) 104032},
\href{http://arxiv.org/abs/1510.05593}{{\ttfamily arXiv:1510.05593 [gr-qc]}}.

\bibitem{Ashtekar:2015ooa}
A.~Ashtekar, B.~Bonga, and A.~Kesavan, ``{Gravitational waves from isolated
  systems: Surprising consequences of a positive cosmological constant},''
  \href{http://dx.doi.org/10.1103/PhysRevLett.116.051101}{{\em Phys. Rev.
  Lett.} {\bfseries 116} no.~5, (2016) 051101},
\href{http://arxiv.org/abs/1510.04990}{{\ttfamily arXiv:1510.04990 [gr-qc]}}.

\bibitem{Ashtekar:2014zfa}
A.~Ashtekar, B.~Bonga, and A.~Kesavan, ``{Asymptotics with a positive
  cosmological constant: I. Basic framework},''
  \href{http://dx.doi.org/10.1088/0264-9381/32/2/025004}{{\em Class. Quant.
  Grav.} {\bfseries 32} no.~2, (2015) 025004},
\href{http://arxiv.org/abs/1409.3816}{{\ttfamily arXiv:1409.3816 [gr-qc]}}.

\bibitem{Teitelboim:1987zz}
C.~Teitelboim and J.~Zanelli, ``{Dimensionally continued topological
  gravitation theory in Hamiltonian form},''
\href{http://dx.doi.org/10.1088/0264-9381/4/4/010}{{\em Class. Quant. Grav.}
  {\bfseries 4} (1987) L125}.

\bibitem{Deser:2011zk}
S.~Deser and J.~Franklin, ``{Canonical Analysis and Stability of
  Lanczos-Lovelock Gravity},''
  \href{http://dx.doi.org/10.1088/0264-9381/29/7/072001}{{\em Class. Quant.
  Grav.} {\bfseries 29} (2012) 072001},
\href{http://arxiv.org/abs/1110.6085}{{\ttfamily arXiv:1110.6085 [gr-qc]}}.

\bibitem{Dadhich:2015ivt}
N.~Dadhich, R.~Durka, N.~Merino, and O.~Miskovic, ``{Dynamical structure of
  Pure Lovelock gravity},''
  \href{http://dx.doi.org/10.1103/PhysRevD.93.064009}{{\em Phys. Rev.}
  {\bfseries D93} no.~6, (2016) 064009},
\href{http://arxiv.org/abs/1511.02541}{{\ttfamily arXiv:1511.02541 [hep-th]}}.

\bibitem{Appelquist:1983vs}
T.~Appelquist and A.~Chodos, ``{The Quantum Dynamics of Kaluza-Klein
  Theories},''
\href{http://dx.doi.org/10.1103/PhysRevD.28.772}{{\em Phys. Rev.} {\bfseries
  D28} (1983) 772}.

\end{thebibliography}\endgroup

\bibliographystyle{./utphys1}
\end{document}